\documentclass[12pt]{spieman}  
\usepackage{amsmath,amsfonts,amssymb}
\usepackage[pdftex]{graphicx}
\usepackage{setspace}
\usepackage{tocloft}
\usepackage{url}
\usepackage[export]{adjustbox}
\usepackage{geometry}
\usepackage{float}
\usepackage{wrapfig,framed}
\usepackage{floatflt}
\usepackage{rotating}
\usepackage{lineno}
\linenumbers

\title{Ultraviolet Type Ia Supernova \textcolor{black}{Mission} (UVIa): Science Motivation \& Mission Concept}

\author[a,*]{Keri Hoadley}
\author[b,**]{Curtis McCully}
\author[c]{Gillian Kyne}
\author[a,d]{Fernando Cruz Aguirre}
\author[b,e]{Moira Andrews}
\author[c]{Christophe Basset}
\author[f]{Azalee Bostroem}
\author[g]{Peter J. Brown}
\author[a]{Greyson Davis}
\author[f]{Erika T. Hamden}
\author[b]{Daniel Harbeck}
\author[c]{John Hennessy}
\author[c]{Michael Hoenk}
\author[h]{Griffin Hosseinzadeh}
\author[b,e]{Andrew Howell}
\author[c]{April Jewell}
\author[i]{Saurabh Jha}
\author[a]{Jessica Li}
\author[f]{Peter Milne}
\author[c]{Leonidas Moustakas}
\author[c]{Shouleh Nikzad}
\author[j]{Craig Pellegrino}
\author[k]{Abigail Polin}
\author[f]{David J. Sand}
\author[l]{Ken J. Shen}
\author[b]{Lisa Storrie-Lombardi}

\affil[a]{University of Florida, Department of Astronomy, Bryant Space Science Center, Gainesville, FL, USA}
\affil[b]{Las Cumbres Observatory, 6740 Cortona Drive, Suite 102, Goleta, CA 93117, USA}
\affil[c]{Jet Propulsion Laboratory, California Institute of Technology, 4800 Oak Grove Drive, Pasadena, CA 91109, USA}
\affil[d]{University of Iowa, Department of Physics \& Astronomy, 203 Van Allen Hall, Iowa City, IA 52242, USA}
\affil[e]{University of California Santa Barbara,}
\affil[f]{Steward Observatory, University of Arizona, 933 North Cherry Avenue, Tucson, AZ 85721, USA}
\affil[g]{Texas A\&M University, Department of Physics and Astronomy, 4242 TAMU, College Station, TX 77843, USA}
\affil[h]{Department of Astronomy \& Astrophysics, University of California, San Diego, 9500 Gilman Drive, MC 0424, La Jolla, CA 92093-0424, USA}
\affil[i]{Department of Physics and Astronomy, Rutgers, the State University of New Jersey, Piscataway, NJ 08854, USA}
\affil[j]{Goddard Space Flight Center, 8800 Greenbelt Rd, Greenbelt, MD 20771, USA}
\affil[k]{Purdue University, Department of Physics and Astronomy, 525 Northwestern Ave, West Lafayette, IN 4790720, USA}
\affil[l]{Department of Astronomy and Theoretical Astrophysics Center, University of California, Berkeley, CA 94720, USA}

\cftpagenumbersoff{figure}
\cftpagenumbersoff{table} 
\begin{document} 
\nolinenumbers

\maketitle

\begin{abstract}
\textcolor{black}{The Ultraviolet (UV) Type Ia Supernova Mission (UVIa) is a CubeSat/SmallSat concept} that stands to test critical space-borne UV technology for future missions like the \textit{Habitable Worlds Observatory} (HWO) while elucidating long-standing questions about the explosion mechanisms of Type Ia supernovae (SNe Ia). UVIa will observe whether any SNe Ia emit excess UV light shortly after explosion to test progenitor/explosion models and provide follow-up over many days to characterize their UV and optical flux variations over time, assembling a comprehensive multi-band UV and optical low-redshift anchor sample for upcoming high-redshift SNe Ia surveys (e.g., Euclid, Vera Rubin Observatory, Nancy Roman Space Telescope). UVIa's mission profile requires it to perform rapid and frequent visits to newly discovered SNe Ia, simultaneously observing each SNe Ia in two UV bands (FUV: 1500 -- 1800 {\AA} and NUV: 1800 -- 2400 {\AA}) and one optical band ($u$-band: 3000 -- 4200 {\AA}). In this study, we describe the UVIa mission concept science motivation and basic mission design. \textcolor{black}{The UVIa mission concept has been submitted to the CubeSats category of the NASA ROSES Astrophysics Research \& Analysis (APRA) program (\$10M cost cap) and NASA Astrophysics Pioneers program (\$20M cost cap).}
\end{abstract}


{\noindent \footnotesize\textbf{*}Prof. Keri Hoadley, \linkable{khoadley@ufl.edu} }

{\noindent \footnotesize\textbf{**}Dr. Curtis McCully, \linkable{cmccully@lco.global} }

\begin{spacing}{2}

\section{Introduction}\label{sec:intro}
\begin{figure}
    \centering
    \vspace{-0.3cm}

    \includegraphics[width=\textwidth,trim=0.0in 1.2in 0.0in 1.2in,clip=true]{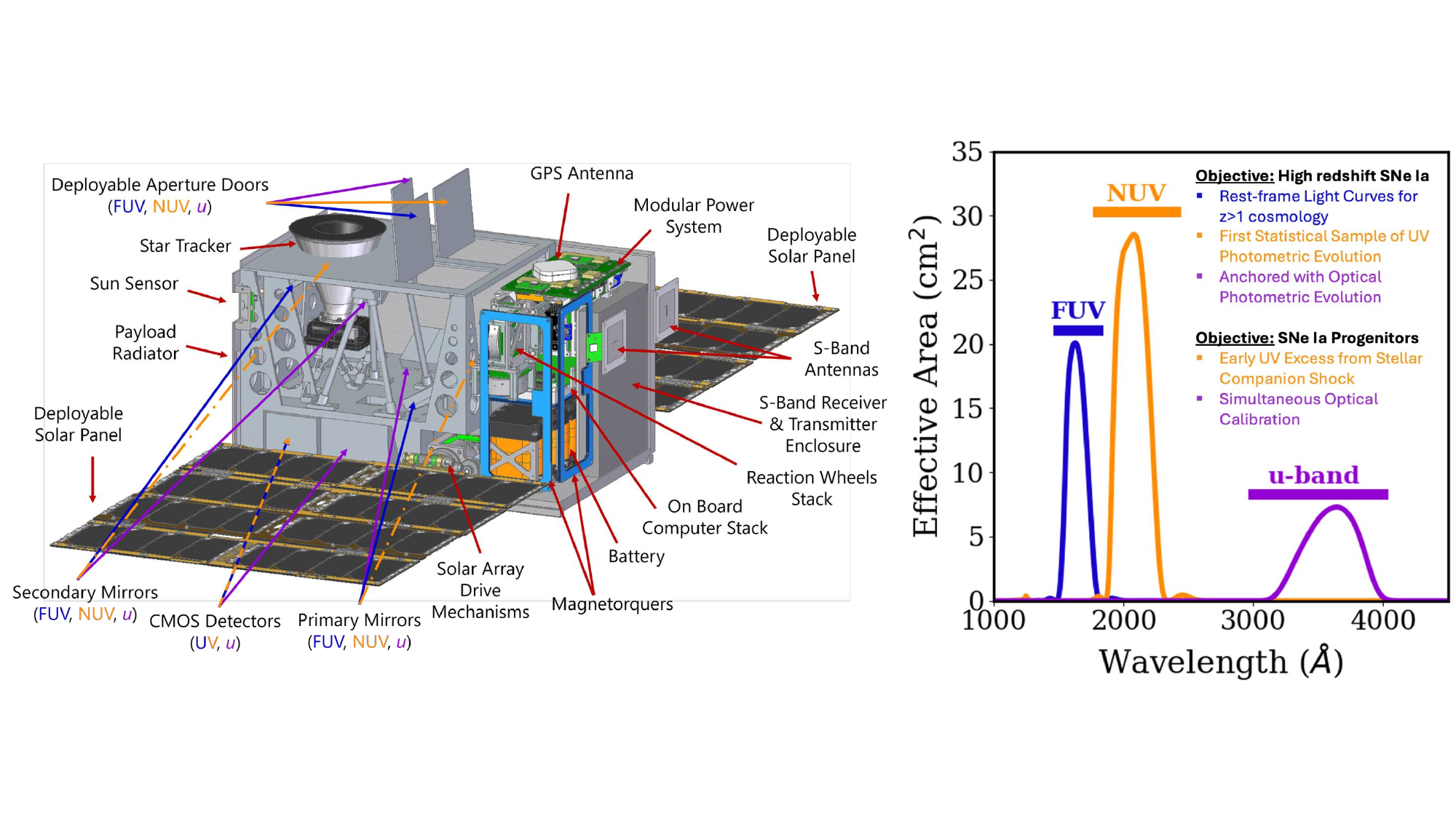}
    
    \caption{\textbf{(Left)}: Model rendering of the UVIa payload, integrated into a commercial off-the-shelf \textcolor{black}{12}-U spacecraft bus (provided by UTIAS-SFL). Dot-dashed lines represent obscured payload components, and the dashed line represents a shared FUV/NUV component. \textbf{(Right)}: Anticipated Effective Area for UVIa's three channels, based on simulations of the optical design and performance. Advances in UV technology and autonomous observatory control make UVIa a viable, efficient observatory to fill in missing gaps about SNe Ia origins and empirical light curves.}
    \label{fig:payload_Aeff}
    
\end{figure}

The Ultraviolet Type Ia Supernova Mission (UVIa) is a Cubesat/Smallsat concept dedicated to rapid and frequent UV/optical follow-up of Type Ia Supernovae (SNe Ia) in a compact, cost-effective \textcolor{black}{($\leq$\$20M)} package (Figure~\ref{fig:payload_Aeff}, left). UVIa is designed as a generalized transient astronomy response platform, to follow up supernova events caught with modern and upcoming ground-based all-sky surveyors (like the Zwicky Transient Factory\cite{ZTF} (ZTF), the Legacy Survey of Space and Time\cite{LSST} (LSST) on the Rubin Observatory\cite{Hambleton_2023}, BlackGEM\cite{Groot+2024}, and Argus\cite{Law+2022}, among others) less than a day after they are discovered. UVIa's prime science sample is SNe Ia and will monitor them for up to 60 days after detection, providing the first statistical UV sampling of SNe Ia light curves out to 100 Mpc. \textcolor{black}{UVIa both compliments and provides a unique perspective on SNe Ia for cosmology and explosion physics within the landscape of current and future UV capabilities: UVIa is designed to observe multi-color UV-to-optical photometry contemporaneously (which UVEX and ULTRASAT lack), while providing high-cadence sampling of the SNe Ia over many days after the initial acquisition (which HST cannot do) with adequate red light suppression to provide unbiased UV photometry (which \emph{Swift} cannot provide). UVIa compliments ULTRASAT's transient discovery objective by offering a reactive, all-sky capability which focuses on a large fraction of SNe Ia within the local cosmological volume, while ULTRASAT adopts a narrow-deep approach, observing more distant objects with limited connection to nearby rungs of the cosmological distance ladder. UVIa would also compliment UVEX's Community Target of Opportunity (ToO) program, providing much-needed constraints on UV light curves to take full advantage of the UV capabilities of the MIDEX mission.}

UVIa contemporaneously observes in three photometric channels: FUV (1500 -- 1800 {\AA}), NUV (1800 -- 2400 {\AA}) and $u$-band (3000 -- 4200 {\AA}). \textcolor{black}{With a total integration time of 1 hour, achieved by stacking 20 images with 300 second exposures in each channel,} UVIa is sensitive to point sources down to an apparent magnitude of 21.5 AB in both UV bands (and 19 mag AB in $u$-band) at a signal-to-noise ratio (S/N) of $\geq$ 5. 
Advances in UV detector\cite{Nikzad+2012,Hoenk+2014,Nikzad+2017,Jewell+2024} and mirror coating\cite{LopezReyes+2021,LopezReyes+2024,Nell+2024} technologies allow UVIa to hit its projected performance in the UV (Figure~\ref{fig:payload_Aeff}, right). UV-sensitive delta-doped Complementary Metal–Oxide–Semiconductor (CMOS) Imaging Sensors (CIS) with integrated metal dielectric filter (MDF) coatings are projected to provide $>$40\% quantum efficiency (QE) in UV bandpasses traditionally only able to achieve $\leq$12\% \cite{Jelinsky+2003}), as well as solar (optical) suppression ($<$10$^{-5}$)\cite{Hoenk+2023,Jewell+2024}. At the same time, emerging advances in multi-layer UV mirror coatings provide a promising avenue to achieve high throughput, narrow-band imaging filters in the FUV and NUV, further rejecting red light that makes accurate UV photometry challenging\cite{Hennessy2017}. 

All UV technologies have been called out as Tier 1 -- 3 priorities in NASA's Astrophysics Technology Gap Priority List\cite{NASATechGap} and support critical advances in UV technology that \textcolor{black}{help make possible sensitive UV instruments, including imagers and photometers, for future UV space-based observatories, including} the Habitable Worlds Observatory (HWO). 
UVIa is designed for a Sun Synchronous (SSO), Low Earth orbit (LEO), and its science operations are executed over the course of 12--18 months. The mission concept is optimized in a compact format, designed such that the payload could fit within a variety of commercially off-the-shelf (COTS) 12U/16U CubeSat or larger spacecraft (s/c) options. 

\section{Science Motivation}\label{sec:science}

Type Ia supernovae (SNe Ia) have historically been used as cosmological distance indicators and led to the discovery of the accelerating expansion of the universe driven by dark energy \cite{Riess98, Perlmutter99}. The longstanding cosmological relevance of SNe Ia has been borne out of the empirical fact that their light curve shapes correlate with their intrinsic luminosity, making them ``standardizable'' candles \cite{Phillips93}. Despite this success of using SNe Ia as cosmological distance indicators, current surveys are already limited by systematics for SNe Ia cosmology\cite{Wood-Vasey07,Kessler09,Conley11}: \textcolor{black}{observing more SNe Ia will not improve cosmological constraints without improving our inference of the distances to SNe Ia. The ``Hubble tension'' -- a 5.9$\sigma $ discrepancy between $H_0$ measured with supernovae ($73.2 \pm 0.9$ km s$^{-1}$ Mpc$^{-1}$) \cite{2024ApJ...973...30B}, and the value predicted by the CMB ($67.4 \pm 0.5$ km s$^{-1}$ Mpc$^{-1}$) \cite{2020A&A...641A...6P}, illustrates that there are either systematic errors in the distance measurements, or new physics is required. Moreover, recent indications (e.g., from DESI\cite{2025arXiv250314738D}) suggest time-varying dark energy whose cosmological signature makes SNe~Ia too bright (by about 0.05 mag) at $z \simeq 0.5$, yet too faint (by about 0.03 mag) at $z \simeq$ 1 -- 2, relative to a cosmological constant model. Rubin and Roman can confirm or reject this model at high significance, but only if we can control for potential SN~Ia evolution over cosmic time. Rest-frame UV observations, sensitive to astrophysical variations in SNe~Ia, are the way to do this, but we currently lack the needed low-redshift sample. The large, high-redshift samples from next-generation surveys like LSST on the Vera Rubin Observatory and the Nancy Grace Roman telescope will not be able to resolve these cosmological ``crises'' without improved light curve modeling in the UV. UVIa will provide this training sample!}

\subsection{The Need for UV: Empirical UV Light Curves for Cosmology}
Constraining UV light curves is key to improving our SN Ia distance measurements. Observationally, the largest scatter in SNe Ia colors (even after \textcolor{black}{matching the stretch for the optical light curves}) is in the UV \cite{Foley16}: HST NUV spectra of a handful of SNe Ia show that the UV has flux variations at the $\sim50\%$ level, even when the optical is nearly indistinguishable \cite{Foley16}. This variation in the UV is not included in cosmological training samples due to the lack of data in the UV with the necessary photometric calibration (e.g. \cite{Guy2007}). 

The dearth of UV calibration samples disproportionately affects our calibration of distance indicators for the highest redshift SN Ia samples, like those that will be observed with the Roman Space Telescope, for which the restframe UV is redshifted into the optical/IR. These require restframe UV light curves to properly calibrate against. Without new UV templates, the conservative approach is to exclude these objects from cosmological analysis to limit the introduction of new systematics. These affect the highest redshift objects: excluding these objects shortens the lever arm in cosmological fits, disproportionately affecting the possible precision on parameters like $w$ that describes the dark energy equation of state. 

\begin{figure}
    \centering
    \includegraphics[width=1.0\textwidth]{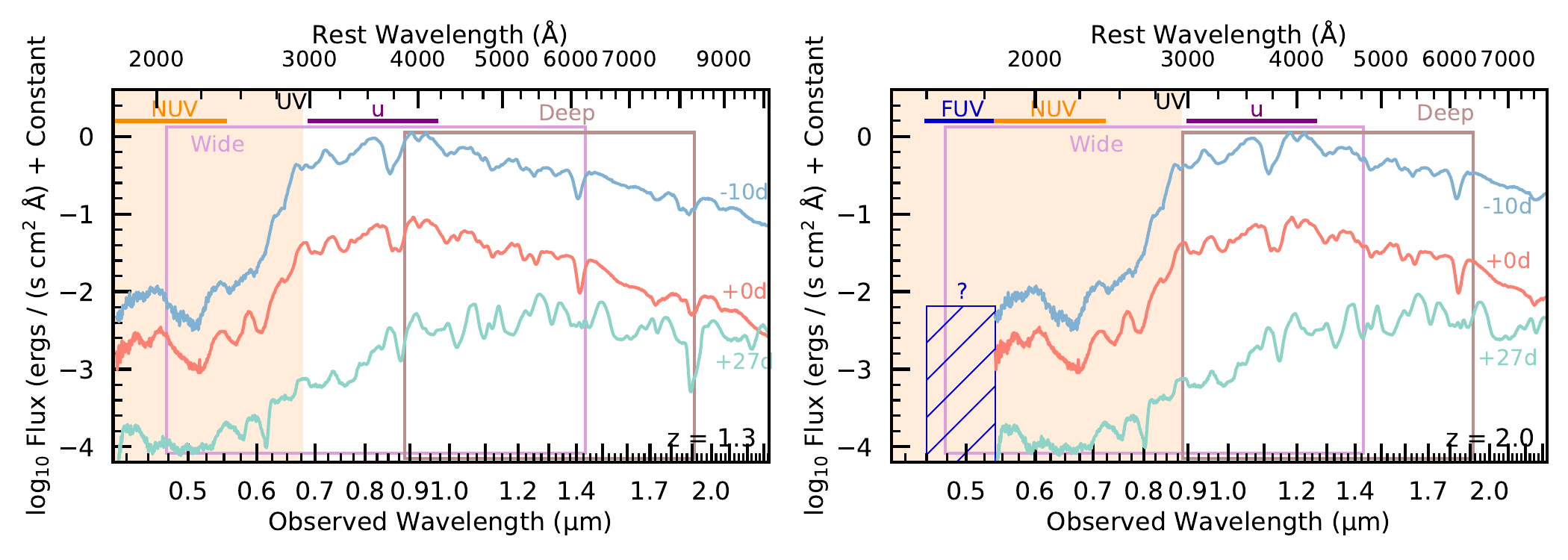}
    \caption{Regions of a SN Ia SED that will be observed by the Roman telescope. Each panel corresponds to a different redshift, $z=1.3$ and $z=2.0$, left and right respectively. These redshifts represent our cutoff where the SN Ia distances will start to be affected by the missing UV calibration data (see text for further discussion). We show boxes that represent the Wide and Deep strategy surveys\cite{Rose21}. The spectra presented here are of SN 2011fe\cite{Mazzali+2014}. UVIa's filter bandpasses are shown near the top of each frame: UVIa will fill this important gap in UV calibration so that Roman \textcolor{black}{can} reach its full potential in SN Ia cosmology.}
    \label{fig:roman-observed}
\end{figure}

As an illustration, we choose a threshold of 20\% of the Roman sample that will be redshifted into the currently uncalibrated UV band to estimate where a mission should include passband to overcome this lack of data. Rose et al.~2021\cite{Rose21} define a fiducial survey strategy for Roman that we then use to estimate the number of objects that will have restframe UV flux in their observations. They employ a two-tier, wide/deep, survey strategy. The bluest filter for the wide survey is $R$ while for the deep survey it is $Y$. Figure \ref{fig:roman-observed} shows the Roman band-passes overlaid with spectra of SN 2011fe at different redshifts.

For the wide survey, 20\% of the SNe Ia are at $z > 1.3$ based on simulations from Rose et al.~2021\cite{Rose21}. The bluest filter on Roman cuts off at 480 nm: at $z = 1.3$, this corresponds to 209nm. Therefore, to fill the need for UV templates, we need observations that go down to about 210 nm so that Roman SNe Ia can be calibrated without extrapolating the templates into the UV.

For the deep survey, 20\% of the SNe Ia are at $z > 2.0$. The blue wavelength limit of $Y$-band, the bluest band for the Deep survey, is 927nm. This corresponds to a restframe of 309 nm which is restframe $u$-band. For the highest redshift SNe, well-calibrated $u$-band data is a necessity for SN Ia templates. For both surveys, 20\% of the sample being unusable without UV observations is non-negligible: \textcolor{black}{simple simulations of the recovered cosmological parameters shows a 50\% increase in the precision on $w_0$ with the full sample compared to the sample excluding the highest redshift objects.}  We need a nearby sample of SNe Ia for UV templates so that Roman can reach its full potential in constraining cosmology.

\subsection{UV Probes of the SN Ia Progenitor Problem}
A significant systematic in SN Ia distance calibration arises due to the lack of detailed understanding of the SN Ia progenitor system. Capturing UV light curves as soon as a few days after a SN Ia explodes is crucial to improving our physical understanding of them. Early-time observations in the UV probe the outermost layers of the ejecta, which is expected to retain the strongest signature for the SN progenitor system. The surface material will be comprised of any accreted material that hasn't undergone further burning, providing a direct probe of the donor star. Depending on the flame propagation speed, this material may also be untouched by the explosion, providing a snapshot of the WD before explosion. Observationally, early-time observations show a diversity of rise-time behaviors in observations $<5$ days after explosion\cite{Ni2024}. These variations disappear shortly after, converging to similar colors and light-curve evolutions. Early observations suggest that there are multiple populations of SNe Ia that would not be differentiated without early follow-up observations\cite{Ni2024}. 

While it is well accepted that a SN Ia is the thermonuclear runaway of a C/O WD\cite{Bloom12}, the nature of the donating companion star and the explosion mechanism are still debated.
There are a variety of mechanisms that produce something that looks like a SN Ia in the optical, like mass transfer from an evolved star like a red giant, mass transfer from a second WD onto the primary either via disruption and then accretion \cite{Iben,Webbink} or by violent merger of two WDs \cite{Pakmor10}. These models often require the primary WD to reach the Chandrashekar Mass and then explode via ``Delayed Detonation'' (DDT) explosion, where the initial burning front begins as a deflagration moving subsonically through the WD until it transitions to a supersonic detonation\cite{DDT}. These DDT models are well tested and match observations outside of the optical in the IR taken with JWST\cite{DerKacy2024}. We will adopt these as the baseline models for the UV properties of SNe Ia, highlighting observational signatures of other models that would favor them over the classic delayed detonation.

DDT models generically predict that the outermost layers of the eject are mostly pristine, untouched by the sub-sonic burning front before it is ejected.\cite{RopkeReview} Iron (Fe)-peak elements absorb most UV photons due to their very high optical depths. As there has been little to no burning in the outer layers, the only source of these elements is the original metallicity of the progenitor. The UV photons are produced by the $\sim10,000 - 15,000$K photosphere, overlaid with absorption of elements in the atmosphere. The UV luminosity rises as the photosphere expands and then declines as the ejecta cools. In addition, as the photosphere reaches more of the material burned during the deflagration, the UV is further absorbed until it is a small fraction of the bolometric flux. This model predicts that the UV flux should rise and fall relatively smoothly as the ejecta is unlayered and well mixed.\cite{HoeflichReview} Abundances of burning products can be estimated by fitting the rise and decline rates to detailed models of the different accretion/merger channels.

Alternatively, pathways for igniting WDs below the Chandrasekhar mass limit have gained popularity in recent years. Models suggest that a sub-Chandra WD that has an outer shell of helium donated from a companion can explode through a pathway known as the ``double detonation mechanism," where first the helium ignites, sending a shock wave into the C/O WD triggering a core detonation followed by a thermonuclear runaway \cite{Nomoto82, Woosley&Weaver94}. These models produce less mixing due to the super-sonic burning front and yield a more layered ejecta structure, if from nothing more than the layer of helium on the surface that detonates to trigger the supernova. If the helium layer is thick, the early light curve can rise more rapidly than the DDT models and will have a redder color\cite{Polin2019} (see SN 2018byg for the discovery of such an object\cite{De2019}). The timescale for the red colors depends directly on the mass of the helium shell.\cite{Shen14HeShell,Polin2019} SN 2022joj showed red colors in the first week after explosion, but by peak was indistinguishable from a normal SN Ia\cite{PadillaGonzalez2024} suggesting small amount of He\cite{Shen14HeShell,Polin2019}. This blue-suppression is a key observable to detect non-negligible helium shells in double detonations. If the UV is suppressed for only a few days, like for SN 2022joj, and then returns to more typical evolution we can infer that the helium shell on the surface was thin. If, however, the SN remains red until maximum or later, we can infer a thick helium shell.  During He shell burning, trace amounts of radioactive material are synthesized (the overall amount will be a function of the initial shell mass). These radioactive elements (primarily $^{48}$Cr and $^{52}$Fe) have very short half lives, which heat the outer layers of the ejecta when they decay, creating flux that escapes over very short diffusion times. This phenomenon has been seen in the optical but is expected to show up earlier and stronger in the UV, where the initial heating peak will quickly lead the way to distinct UV colors \cite{Magee2021}. The strength of this heating and excess UV flux is related to the thickness of the helium shell that detonates. Recently, a new class of models have been suggested: dynamically driven, double detonation of double degenerate system (D6) models\cite{Shen21} has been proposed as a mechanism with negligible He on the surface of the WD at time of explosion. The minimal amount of helium leads to no significant production of these short-lived isotopes and no excess UV heating. While the D6 models work exceedingly well in the optical, these models underpredict the UV flux at early phases for SN 2011fe (the models match at peak).\cite{Shen21_nonlte}. The UV is again a distinguishing waveband for explosion models, in this case, thick helium shell double detonations and the D6 models.  

In addition to the explosion mechanism, the nature of the donating companion star to these WD explosions is still under debate. Traditionally, there are two classes of companion star models for SN Ia progenitors, the ``Single Degenerate'' model (where a non-degenerate red giant transfers mass to the WD via Roche-lobe overflow \cite{Whelan}), and the ``Double Degenerate'' models that have two WDs that merge either by disruption and accretion of the smaller onto the larger WD\cite{Iben,Webbink} or by violent merger\cite{Pakmor10}. The presence of a red giant companion star has a predicted signature for the Single Degenerate model: for favorable viewing angles, there should be an excess of UV emission in early-time observations, as the ejecta is shocked when it collides with the companion star \cite{Kasen10, Brown12Companion}. If our line of sight is down the ``hole'' in the ejecta produced by the companion star blocking part of the ejecta from expanding, we expect to see excess emission in the UV, while for other viewing angles, the rest of the ejecta will hide some or all of this excess emission. Based on the binary separation and the radius of the companion, we should see this effect in $<10\%$ of single degenerate SNe Ia\cite{Kasen10}. This UV excess is expected to have a higher temperature ($\sim 3\times 10^5 K$) than radioactive heating and the shock cooling follows a power law decay\cite{Kasen10} rather than radioactive decay like the excess predicted for short lived isotopes ($^{48}$Cr and $^{52}$Fe) produced in double detonations. UV colors are essential to characterize the temperature evolution to distinguish between companion shocks and additional radioactive power in the explosion itself.

\begin{figure}
    \centering
    \includegraphics[width=\textwidth]{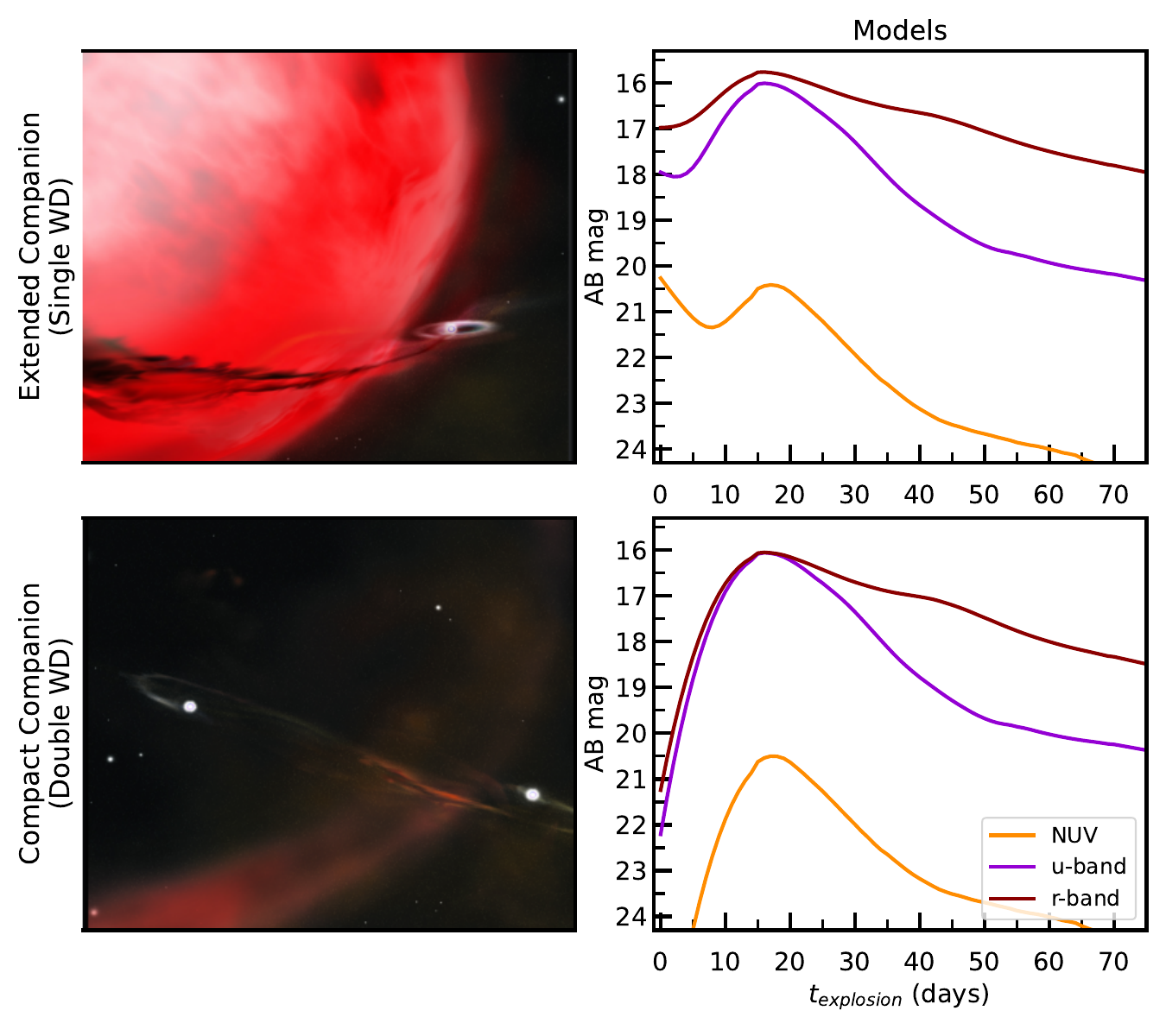}
    \caption{Illustration of the differences in the expected light curves with and without companion interaction\cite{Kasen10}. Image credit: NASA Goddard Space Flight Center. For the companion model on the top, we have adopted a 1 $M_\odot$ red giant (binary separation $2\times10^{13}$ cm). The effects of the ejecta colliding with the companion are strongest in the blue, necessitating UV observations, and the effects fade quickly requiring rapid response time. These are both features of UVIa's design.}
    \label{fig:companion-illustration}
\end{figure}

This companion shocking effect has not \textcolor{black}{been} observed at the levels predicted for the classic single degenerate model, ruling out red giant companions for the majority of SN Ia progenitors\cite{Bianco11, Bloom12}. There are, however, some cases of an early UV excess similar to what was predicted by Kasen 2010\cite{Kasen10}. SN 2017cbv is one of the best observed SN Ia with an early UV excess\cite{Hosseinzadeh17}. Rather than the excess we would have expected from a red giant, SN 2017cbv's UV excess was more in line with a 7$M_\odot$ main sequence star. Figure \ref{fig:companion-illustration} illustrates the excess seen from a companion like that of SN 2017cbv. A similar excess with comparable size donor star was also found in SN 2019yvq\cite{Burke21}. This implies that there must be at least some SN Ia that arise from binaries that have a companion that is substantially larger than a WD. The binary companion for such a sub-Chandraskhar mass explosion is the variable which determines the mass of helium donated to the primary WD before explosion. A larger companion, like a He-sub dwarf or the stripped core of the companion star (after multiple rounds of mass transfer\cite{Wong24}) could donate a large amount of helium before detonating, producing a double detonation. This He-star companion may not be compact like a normal C/O WD and could produce a UV excess like those predicted for red giants or main sequence stars. Observations in the UV within days of explosion provides a window to measure the otherwise hidden donor star in SNe Ia.

One almost completely unexplored regime to constrain SN Ia progenitors is the FUV. Currently, FUV observations only exist for a \textit{single} SN Ia (SN 2011fe), \textcolor{black}{taken with HST}\cite{Mazzali+2014}. \textcolor{black}{No other instrument with FUV coverage has had the ability to respond rapidly enough to observe SNe.} UVIa FUV observations will provide crucial constraints on present-day models of SNe Ia: the FUV is extremely sensitive to the outermost layers of the progenitor system. Many of these models converge in the optical by maximum light making them indistinguishable. FUV observations can break this degeneracy distinguishing between the proposed explosion mechanisms providing sorely needed constraints. Combined FUV/NUV color observations will provide unprecedented measurements of the color evolution of SNe Ia for the full population and will be essential to distinguish between the physical processes that produce UV excesses.

In summary, UVIa's observations will provide:
\begin{itemize}
     \item FUV -- optical templates used as a legacy dataset to train SN cosmology models which disproportionately affect high redshift surveys like Roman
     \item A statistical sample in the UV, the wavelength band with the largest observed scatter, to empirically characterize the diversity of SNe Ia
     \item NUV and FUV observations to fit to Chandrashekar delayed detonation scenarios to differentiate explosion models based on the isotopes in the outermost layers of the ejecta
     \item Population fraction that shows a suppression of the UV likely driven by a thick shell of helium in a double detonation
     \item Population fraction with a UV excess with colors showing cooling that follows radioactive decay from short lived isotopes like $^{48}$Cr and $^{52}$Fe which are produced in a double detonation
     \item Population fraction with a UV excess driven by shocking from the companion star which is distinguishable using the UV colors that show a power-law decline in the temperature
     \item The first statistical sample of FUV light curves of SNe Ia to be used to constrain explosion models in a completely unexplored frontier
 \end{itemize}

\subsection{Sample Size}
For UVIa to probe systematics in SN Ia distance calibration, we need a statistical sample of SNe Ia where the counting statistics are comparable or better than the current intrinsic scatter in the inferred SN Ia distances. To estimate this intrinsic scatter and in turn, the number of SNe needed, we fit a light curve stretch model \cite{Tripp} to a large sample of low-z SNe Ia from Foley et al.~2018\cite{Foley18}. Our model is
\begin{equation}
DM = m + M_{\rm{SN}} + \alpha x_1 + \beta c 
\end{equation}
where $DM$ is the distance modulus, $m$ is the measured peak magnitude, $M_{\rm{SN}}$ is the intrinsic absolute magnitude of SNe Ia, $x_1$ is the light curve stretch, and $c$ is color term (see Guy et al.~2007\cite{Guy2007}). $\alpha$, $\beta$, and $M_{\rm{SN}}$ are free parameters that are fit to the sample. We include an intrinsic scatter term in our likelihood following Hogg et al.~2010\cite{Hogg10}. We marginalize to estimate the posterior probability distribution on the Hubble constant using Emcee\cite{emcee}. We find that the intrinsic scatter remaining in the SN Ia sample is 0.15 mag. To be able to characterize the full variation in the population, our statistical uncertainties need to be at a similar level to this. To achieve a comparable, fractional Poisson uncertainty requires $\approx50$ SNe Ia, which we then adopt as our sample size requirement for our cosmological analysis.

Early models of companion interaction suggest that viewing angle effects will limit UV bumps in the lightcurve to only be visible in 10\% of cases, even if SD models dominate the SNe Ia progenitor population\cite{Kasen10}. To have a high probability of having at least one UV excess, we need $\approx50$ SNe Ia ($\approx 5$ with an excess if all SNe Ia are single degenerate). Using simple counting statistics, if we detect zero UV excesses, we can rule out that the single-degenerate channel is not dominant ($<43\%$ of the population to 90\% confidence). If we detect 5, the single-degenerate mechanism must be the primary channel, accounting for $>50\%$ at 90\% confidence. Measuring a number in between will constrain the single-degenerate fraction to $\approx25\%$. \textbf{UVIa's science requirement is to produce NUV-u band lightcurves of $>50$ SNe Ia.}

In the FUV, there is an extreme dearth of observations. There is a single spectrum of a single SN Ia (SN 2011fe\cite{Mazzali+2014}). To be able to characterize the population of SNe Ia, we need to go beyond a single object to a statistical sample. To begin to characterize the FUV (including evolution) across the population, we aim to increase the sample of objects with FUV by an order of magnitude for the only one that currently exists. \textbf{UVIa's next science requirement is to produce $>10$ FUV lightcurves of SNe Ia.}

These light curves and the upper limits for the rest of the NUV sample will be included in the cosmology training sets and provide modelers with constraints on a previously completely open part of parameter space. 

\textcolor{black}{Despite many years of effort, the dominant explosion mechanism is still hotly debated. With a sample size of 50, UVIa would provide statistically significant measurement of the fraction of the dominant UV production channel. Sub-dominant channels may only provide upper limits, but even that would move the field forward. For individual SNe Ia that match the expected properties for less common explosion mechanisms, UVIa would perform detailed studies on the specific objects providing additional constraints on those explosion models.}

\subsection{SN Ia Rate}\label{sec:rates}
To ensure we meet this sample size requirement, we next need to estimate the number of SNe Ia that will be observed over the lifetime of the mission. We adopt the rates of Perley et al.~2020\cite{Perley20} as our fiducial intrinsic SN Ia rate, 23,500 SNe Ia / Gpc$^3$ / year. If we assume that the UVIa's field of regard is half of the sky at any given time, we estimate that UVIa will be able to follow 15 SNe Ia within 60 Mpc and 86 within 100 Mpc over the 18-month observing mission lifetime. 

This is comparable to the actual discovery/classification rate of Perley et al.~2020\cite{Perley20}, who discovered and spectroscopically classified $56$ SNe Ia within 100 Mpc during their 25.5 month-long study. Their targets were exclusively drawn from $>-30^\circ$ declinations. We will draw candidates from the entire sky (leveraging other discovery programs than ZTF alone) giving UVIa an additional 33\% sky coverage over the results presented by Perley et al.~\cite{Perley20}. Correcting from their mission lifetime of 25.5 on-sky months to our 18-month mission and adding in the additional sky coverage suggests that UVIa would observe 52 SNe Ia. At this magnitude, ZTF was only 80\% complete due to their stringent cuts for their sample to determine supernova rates which requires a sample with extremely high purity. This yields another $\sim10$ objects, increasing the sample of UVIa follow-up objects to $\sim$62. 
We will also be drawing targets from multiple surveys, leading to better redundancy against weather and mechanical issues.
We conclude that with the current and upcoming discovery surveys and a dedicated follow-up program to classify all SNe discovered within 100 Mpc, UVIa can robustly, with margin, observe $>50$ SNe Ia during its mission. 

\subsection{SN Ia Brightness Estimates}
SN 2011fe is one of the nearest SN Ia in decades which led to it being the best observed SN Ia in all passbands. SN 2011fe was a prototypical SN Ia near the center of the luminosity distribution of SNe Ia \cite{Zhang16}, so we adopt it as a fiducial spectral energy template for our estimates.

To estimate our required depths, we shift SN 2011fe's SED to 100 Mpc. We find that a SN Ia at 100 Mpc will be 21.5 AB mag in the NUV at a phase of $-7$d (relative to peak in the $B-band$). Observations at this depth will be able to cover the NUV light curve to +20 days after peak, after which the UV contribution to the bolometric luminosity is minimal.  In the u-band, we require more complete sampling of the light curve, out to +40 days past maximum. This yields a required depth of 19 mag AB. This depth corresponds to -12 days before peak, providing full coverage of the evolution of the SN. 

As no observations in the FUV exist for SN Ia at early time, we \textcolor{black}{adopt} observations of SN 2011fe in the NUV\cite{Pereira13}, extrapolating the continuum into our passband. Due to faintness in the FUV, we consider SNe Ia at 80 Mpc which would still include $\approx45$ objects. At 80 Mpc, our anticipated FUV brightness a week before maximum is 21.5 mag AB. At later times (e.g., peak brightness), there is evidence that the FUV flux relative to the NUV drops considerably \cite{Mazzali+2014}, but the overall brightness of the SN increases, counteracting some of this dropoff. Due to this uncertainty, we use this image depth to conservatively meet our goal of 10 objects in the FUV, still successfully increasing the current sample by an order of magnitude.

\subsection{Observational Requirements}
Given these depth requirements, we can define the other salient observational parameters that will be necessary to achieve the necessary light-curve precision for the UVIa SNe Ia sample. 

In the NUV and FUV, we adopt a signal-to-noise (S/N) requirement of $>5$ at our required depths to distinguish the light curve evolution in these bands. In the u-band, we require higher S/N of $>10$ as this band is so crucial for high-redshift cosmological calibration. 

The angular resolution is set such that SNe Ia targeted by UVIa are distinguishable from other contaminating sources; a spatial resolution $<$10$^{\prime \prime}$ closely matches the capability of the All-Sky Automated Survey for Supernovae (ASAS-SN) survey, which successfully disentangles transient object photometry from other sources\cite{Kochanek+2017}. UVIa plans to implement their methodology in the data reduction and analysis pipeline to retrieve and measure daily UV and $u$-band photometry for nearby SNe Ia. The FOV requirement of $>$1$^{\circ}$ ensures that a minimum of 3 calibration stars are present in each field per channel to monitor the variation in brightness of SNe Ia over subsequent passes and provide a benchmark for photometric variation between observing fields on different days. While photometric calibration for each channel will happen separately and regularly throughout the mission profile, a large FOV is required to assess relative changes to SNe Ia brightness relative to more stable astronomical sources over time.

UVIa must be reactive to capture the full evolution for use in cosmological calibration and to enable the detection of UV excesses. We require that updated schedules with newly discovered targets be uploaded at least twice daily. We will not wait for classification of newly discovered SNe and will discontinue objects if they do not meet our criterion. This ensures that we have the earliest observations possible. 
UVIa will employ a 1-day cadence to fully sample the light curves. If we find 86 targets over the 18-month, we should be following $\sim5$ objects per month, and observing them for 2 months, we expect to be observing $\sim10$ objects per day. If we assume an hour exposure time, this will correspond to $\sim14$ out of the 16 orbits available to UVIa per day. The final orbits will be used for margin and to follow-up newly discovered SNe before they are classified. Our baseline cadence samples the time scale of the light curve evolution and optimizes the available observing time spent on target.

\begin{figure}
\includegraphics[width=\textwidth]{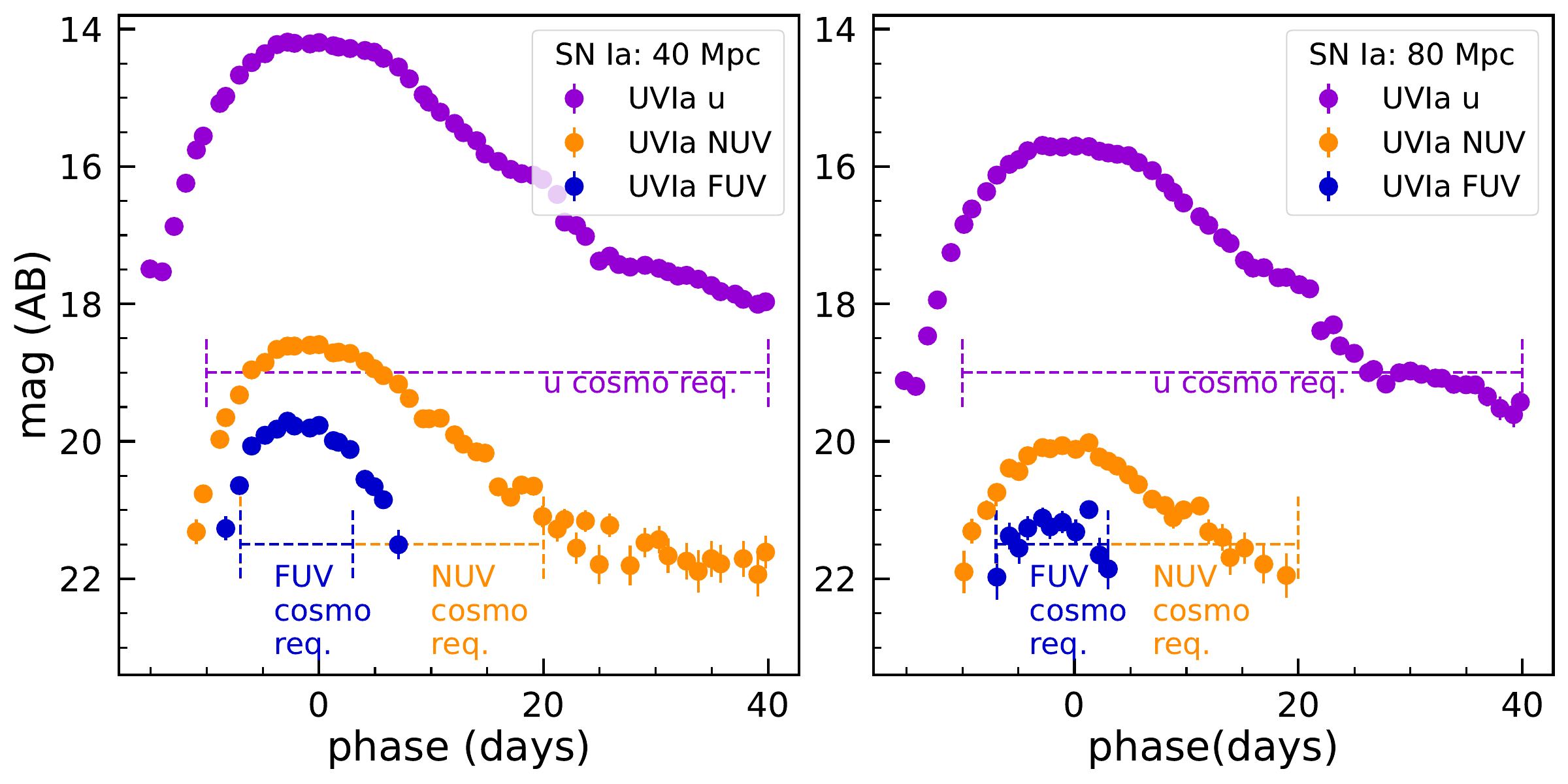}
\caption{Simulated expected $u$-band -- FUV light curves of a SN Ia at multiple distances. We have adopted SN 2011fe as a template SN Ia\cite{Brown1211fe, Zhang16}. UVIa's requirements will allow us to produce $u$-band and NUV light curves to use as a cosmological training set. In the FUV, we will produce high-quality light curves for an order of magnitude more SNe Ia than current samples. We will also be able to use the FUV light curves to constrain the SED in the cosmological training sample.}
\end{figure}

\begin{figure}
\includegraphics[width=\textwidth]{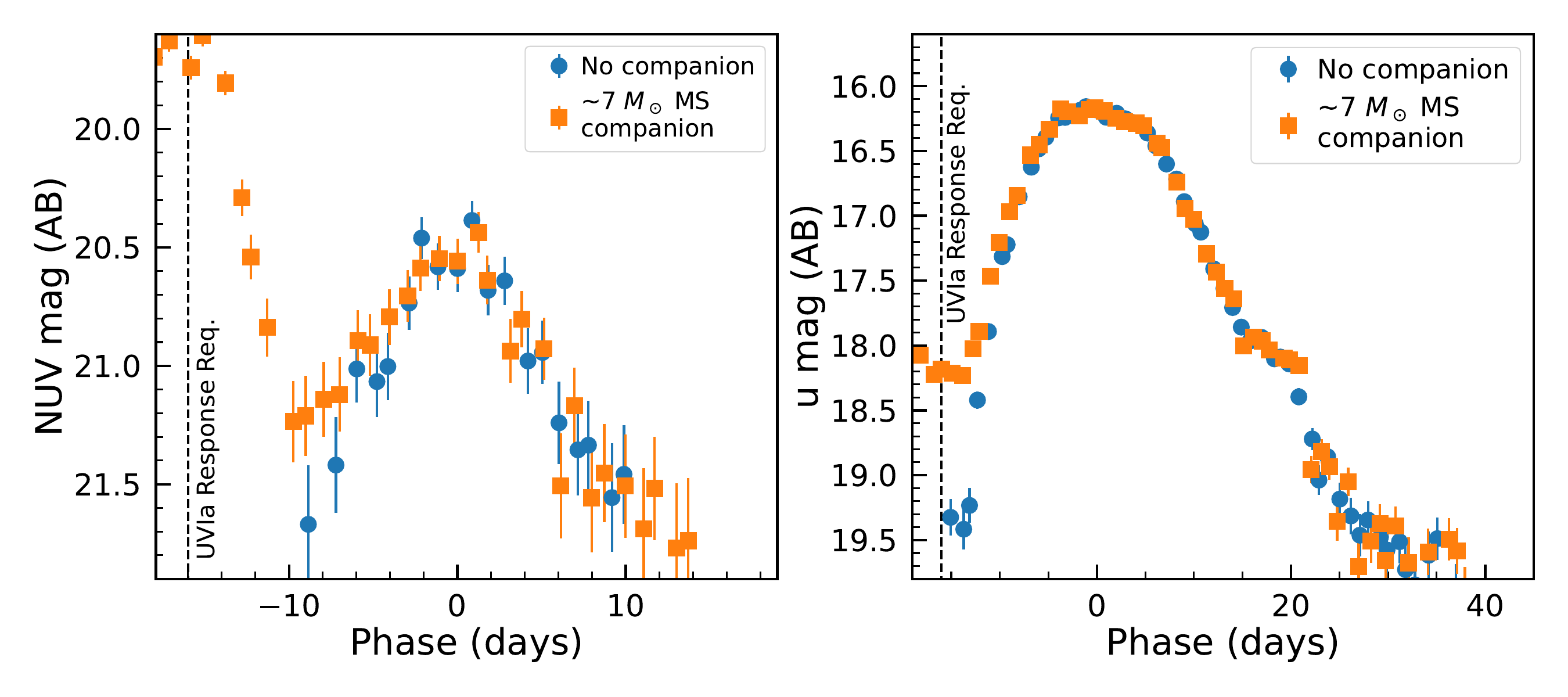}
\caption{Expected $u$-band and NUV lightcurves for a SN Ia given the science requirements of UVIa. We adopt the binary separation of $3\times10^{12}$~cm, which is similar to both SN~2017cbv\cite{Hosseinzadeh17} and SN 2019yvq\cite{Burke21}, added to the light curve of SN 2011fe\cite{Brown1211fe, Zhang16}. This binary separation corresponds to a $\sim 7 M_\odot$ main sequence companion\cite{Kasen10}. Given UVIa's sensitivity and response requirements, we will be able to detect the presence a UV excess like that of SN 2017cbv.} 
\end{figure}

\subsection{Ancillary Science Targets}
Because we trigger on every young supernova with 100 Mpc, we will have early FUV and NUV observations of a variety of classes of explosions. Early UV observations of all types of SNe are rare and have the potential to advance several open questions like the progenitors of exotic thermonuclear SNe (e.g. SNe Iax which are hotter than normal SNe Ia at early times\cite{JhaIaxReview}), interaction with circumstellar material (e.g. type IIn SNe), progenitor constraints on massive stars from shock cooling as the early emission goes from the UV into the optical\cite{Brown07}, and the characterization of the zoo of fast transients (e.g. Fast Blue Optical Transients; FBOTs\cite{Drout14}). If the early observations are novel and can be used to probe one of these physical regimes, we will leave the target in the schedule as a filler target with low priority, allowing us to fill gaps in the schedule and maximize the scientific return of UVIa.   

\section{The UVIa Mission Profile}\label{sec:mission}
The UVIa mission profile (including instrument, payload, and spacecraft requirements) is designed such that its core science observation needs are met, while ensuring a compact design compatible with a minimum of a 12U CubeSat platform. In this section, we briefly describe the UVIa telescope design, new advances in UV filters and coatings that create UVIa’s UV bandpasses, anticipated performance based on conservative estimates of SNe Ia signal and noise sources, resulting spacecraft requirements needed to make a mission like UVIa possible, and projected science operations.

\subsection{Instrument Overview}
\begin{figure}
    \centering
    \vspace{-0.3cm}

    \includegraphics[width=0.95\textwidth]{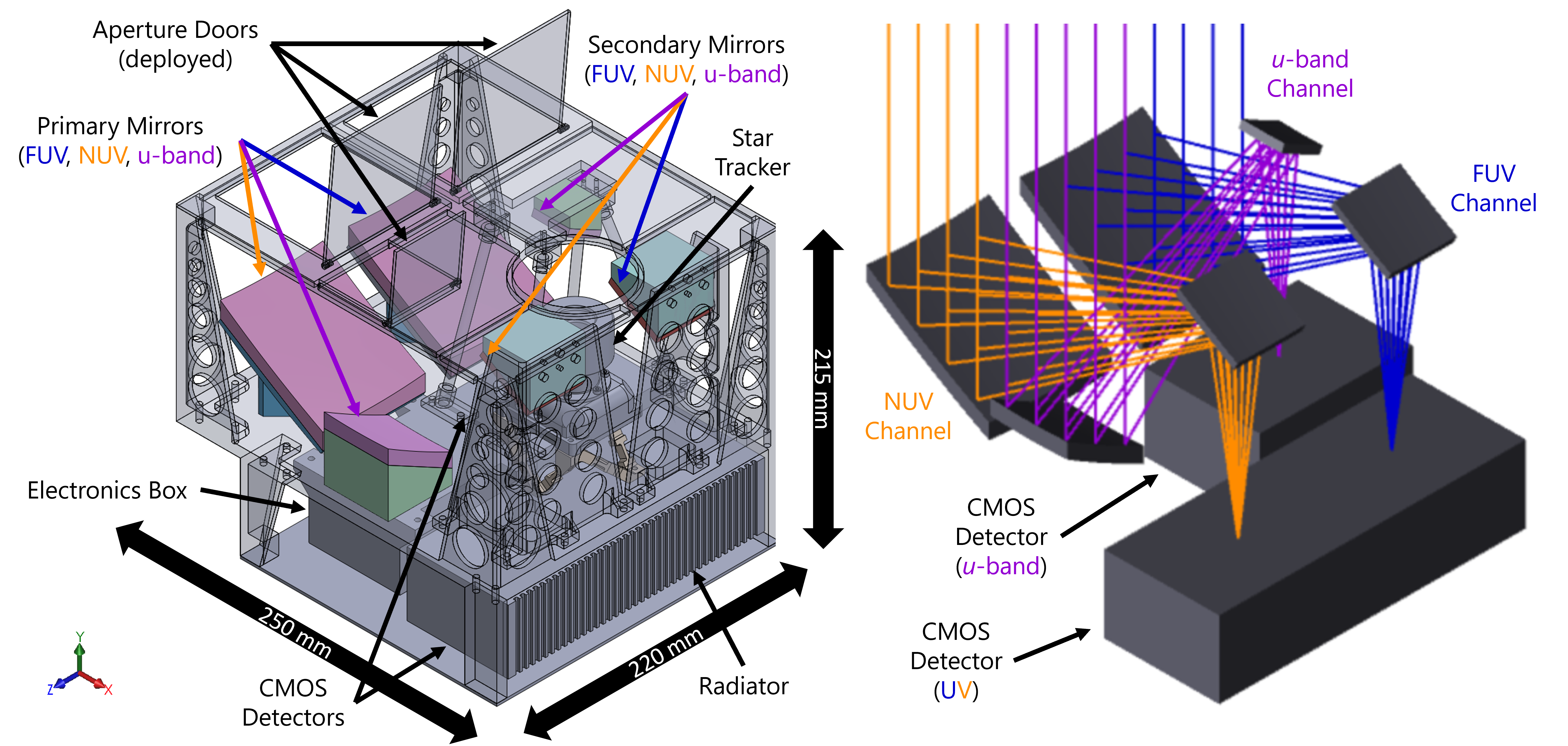}
    
    \caption{\textbf{(Left)}: Model rendering of the UVIa payload, integrated into a commercial off-the-shelf \textcolor{black}{12/16U} spacecraft bus (credit: UTIAS-SFL). \textbf{(Right)}: \textcolor{black}{Ray trace of the FUV (blue), NUV (orange), and $u$-band (purple) channels. Each channel focuses light onto a CMOS sensor. The UV CMOS detector has two active areas.}}
    \label{fig:payload_raytrace}
    
\end{figure}

The UVIa scientific instrument (Figure~\ref{fig:payload_raytrace}) consists of 3 imaging telescopes that cover photometric bands between 1500 -- 4500 {\AA} (FUV: 1500 -- 1800 {\AA}, NUV: 1800 -- 2400 {\AA}, $u$-band: 3000 -- 4500 {\AA}). 
A ray trace of the UVIa telescopes is shown in Figure~\ref{fig:payload_raytrace}. All UVIa telescopes are double-offset Cassegrain telescopes by design, driven to fit within a compact payload volume for a 12U CubeSat while ensuring quality on-axis spatial resolution with a fairly wide field of view (FOV)\cite{CruzAguirreJATIS}. Each UV channel mirror is coated with a multi-layer (ML) thin film coating of alternating aluminum fluoride and lanthanum flouride (AlF$_3$/LaF$_3$), deposited via atomic layer deposition (ALD), where the number of fluoride layers and thicknesses determine each UV bandpass\cite{Larruquert+2011,Larruquert+2022,LopezReyes+2024}. The $u$-band channel mirrors baseline a standard aluminum (Al) with a thin magnesium fluoride (MgF$_2$) over-coating and the SLOAN $u$-band transmission filter. We note that, should the Rubin/LSST $u$-band filter be commercially available, UVIa will use that in place of the SLOAN filter. in the optical path in front of the optical camera. All channels feed to their own Teledyne-e2v CIS120-10-LN (``Capella'') CMOS sensor, featuring a 2k x 2k focal plane with 10$\mu$m pixels and low intrinsic noise ($\leq$7 $e^{-}$/pix read noise and $<$0.05 $e^{-}$/pix/s dark current at an operating temperature T$\sim$-35$^{\circ}$C). The $u$-band CMOS is back-illuminated with UV-enhancement, with projected efficiency based on performance reported for the Ozone Mapping and Profiler Suite (OMPS) mission\cite{Jaross+2014}. Each FUV and NUV CMOS is back-illuminated and delta-doped\cite{Hoenk+1992, Nikzad+2012} with a custom integrated metal dielectric filter (MDF) made up of 7 layers of Al and AlF$_3$; both UV CMOS follow processes developed of the Star-Planet Activity Research CubeSat (SPARCS) to optimize UV efficiency in each FUV and NUV bandpass while suppressing out-of-band light detection\cite{Jewell+2024}. 
For further details on the optical design\textcolor{black}{, trade spaces considered,} and technology associated with UVIa's custom UV mirror coatings and detector filters, see Ref. \citenum{CruzAguirreJATIS}.

All telescopes have a FOV of 6.25 degrees$^{2}$ on the sky, and the on-axis point spread function (PSF) of each telescope channel produces diffraction-limited angular resolution ($<$1$^{\prime \prime}$ in both UV channels and 2.4$^{\prime \prime}$ in the $u$-band channel). This PSF is much smaller than the plate scale at each UVIa camera (6$^{\prime \prime}$/pixel).

\subsection{Anticipated Performance}\label{sec:performance}
\begin{figure}
    \centering
    \vspace{-0.5cm}
    \includegraphics[width=\textwidth,trim=4.1cm 0.0cm 3.5cm 0.5cm,clip=true]{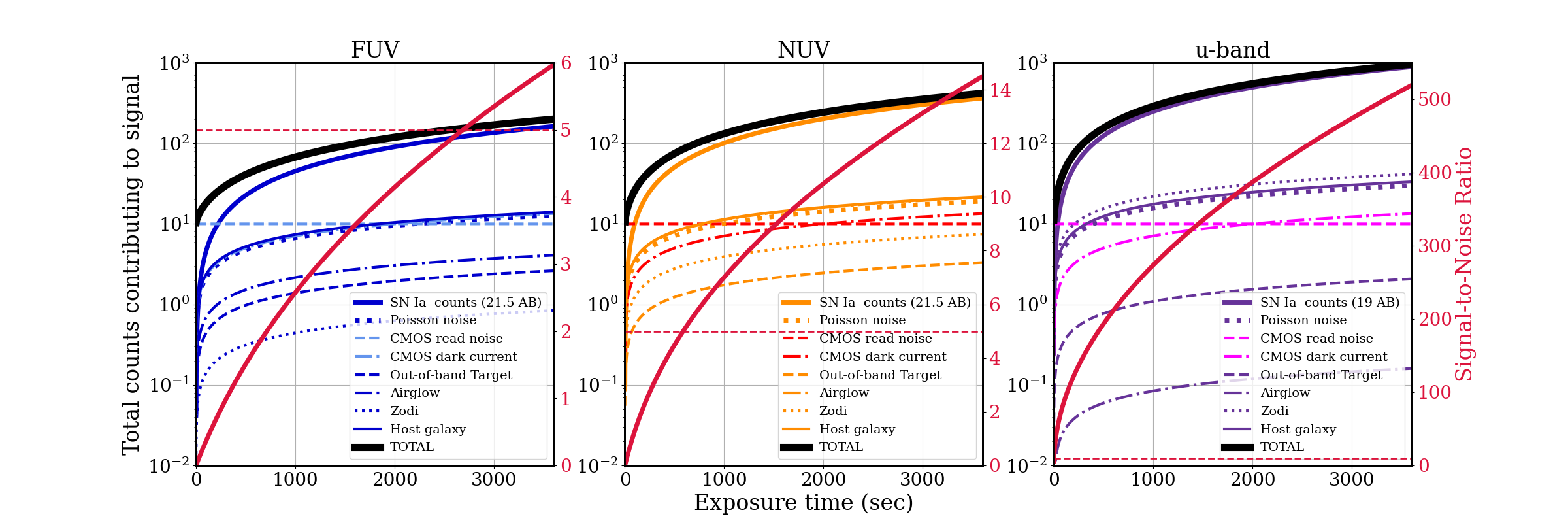}
    \caption{Based on the \textcolor{black}{estimated total counts (as a representation of photons) as a function of exposure time in each UVIa bandpass}, along with worst-case estimates of a variety of noise and contamination sources, we demonstrate that the sensitivity of a transient/SNe Ia at UVIa's minimum point source detection (21.5 AB mag in the FUV/NUV and 19 AB mag in $u$-band) is well above all sources of noise \textcolor{black}{and that the signal-to-noise ratio in each channel meets the minimum needed for the UVIa science objectives in a total of 1 hour of exposure time.} In the FUV, detector noise (RN, DC) and the galactic host of the SNe dominate the noise contributions. In the NUV, the CMOS RN and galactic host contamination are the \textcolor{black}{dominant} factors in the noise. In $u$-band, zodiacal scattered light and galactic host are both primary noise sources. At the required integration time, the signal counts outweigh the noise counts in all bands by a substantial margin.}
    
    \label{fig:count_breakdown}
\end{figure}

UVIa's projected performance is based on the presented optical design and modeled/measured efficiencies of all UV and optical components integrated into the telescope design. The projected performance takes into account the in-band and out-of-band signal from the source (SN Ia) and noise contributions from the detector (read noise (RN), dark current noise (DC)), and in-band and out-of-band contributions from stray light (zodiacal and interstellar scattered light), sky background (Earth UV airglow features), and host galaxy contamination. Each noise contribution includes additional margin on conservative laboratory measurements (when available) and simulations of astronomical and stray light sources, based on empirical data from previous/current facilities. \textcolor{black}{The projected counts as a function of source includes red leak contributions where applicable (e.g., out-of-band signal sources) and jitter effects from the s/c (taken into account as smearing of the target signal over multiple pixels; see Ref. \citenum{CruzAguirreJATIS} for more details and the optical error budget, which takes into account anticipated effects of s/c jitter).}
Figure~\ref{fig:count_breakdown} shows the projected breakdown of total counts from each source contributing to the signal pixel. Below, we describe how the expected count rates from each source are tabulated. 

\textbf{SNe Ia in-band signal:} To date, only one Type Ia supernova, SN2011fe, has a complete, observed FUV-through-optical spectrum\cite{Mazzali+2014}. We use this one example as a ``template'' spectral energy distribution (SED) and scale the measured flux to UVIa's desired minimum sensitivity (21.5 mag (AB) through the UV and 19.0 mag (AB) in $u$-band) to predict the in-band count rates for UVIa's faintest target sampling. This constitutes the anticipated \textit{minimum} in-band science signal count rate per pixel in each UVIa channel. \textcolor{black}{Total signal counts} in all bands as a function of time are shown in Figure~\ref{fig:count_breakdown}. It is worth noting that the only FUV coverage of a SNe Ia to-date was observed about 2 weeks after the initial detection of the supernova, 3--4 days after peak $B$-band brightness\cite{Mazzali+2014}. This is likely not representative of early-time UV detections UVIa would make, but it gives a conservative estimate of the UV fluxes and SED slopes.

\begin{figure}
    \vspace{-0.3cm}
    \centering
    \includegraphics[width=1.0\textwidth]{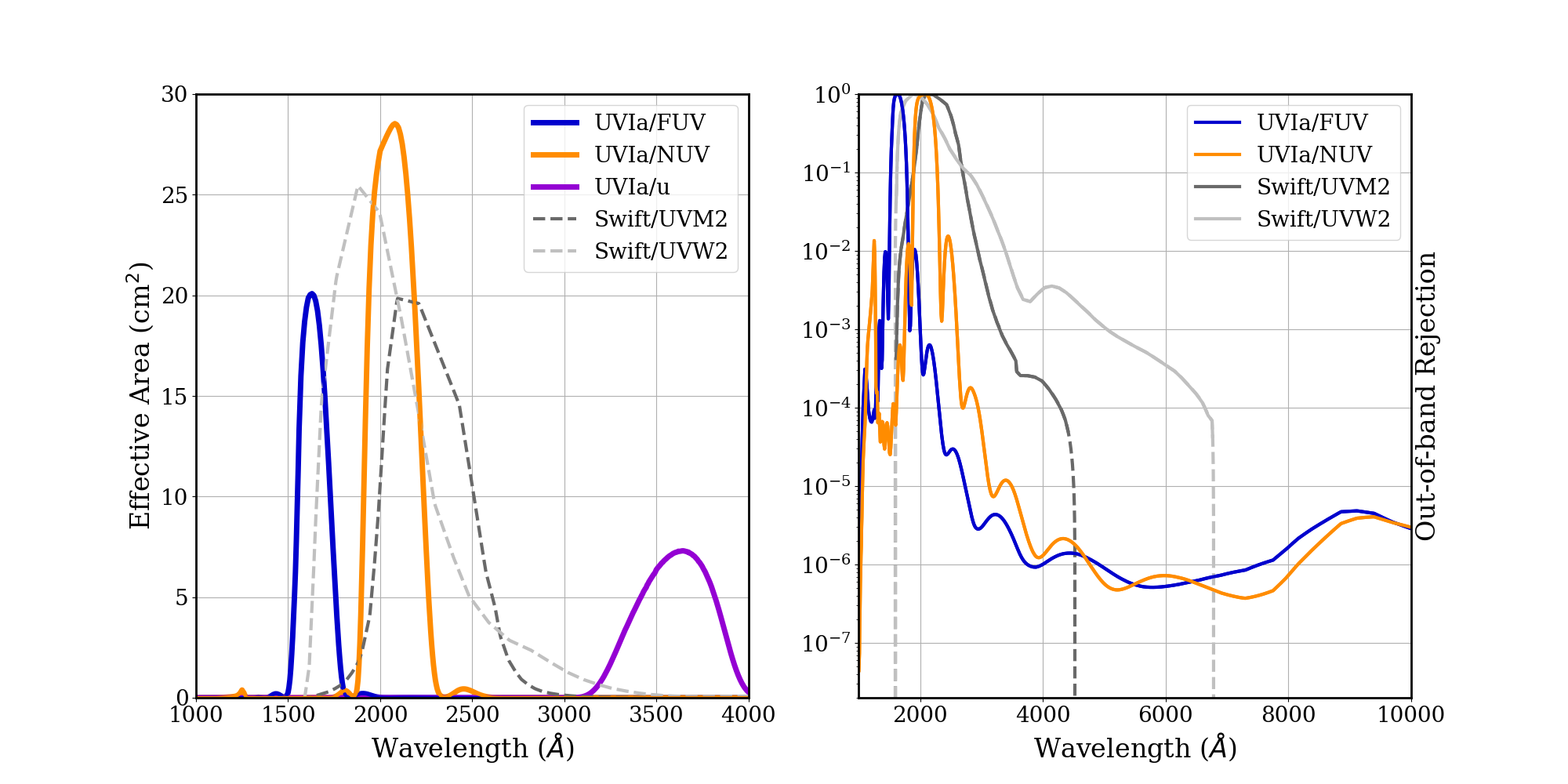}
    
    \caption{UVIa's projected performance (effective area: left and out-of-band suppression: right) in the UV compared to \textit{Swift}-UVOT UVW2 and UVM2 bands. UVIa's channels are shown as solid blue (FUV), orange (NUV) and purple ($u$-band) lines, and the \textit{Swift}-UVOT filters are shown as dashed/solid lines in dark gray (UVW2) and light gray (UVM2). The two \textit{Swift} filters were chosen because they are the two bands that most closely match UVIa's FUV and NUV bandpasses for comparison. \textit{Swift}/UVOT performance is reported in \cite{Breeveld+2010}, where effective areas are only reported to a cutoff point in the optical. For a conservative comparison of UVIa to \textit{Swift}'s current UV capabilities, we assume the \textit{Swift} UV band effective areas fall to 0 cm$^2$ where not reported. Despite being a smaller experiment, UVIa's UV channels are not only comparable to \textit{Swift} in effective area, but achieve narrower UV bands overall with $>$2 orders of magnitude less sensitivity to photons $>$3000{\AA}. This combination allows UVIa to perform critical time-domain science, particularly supernova observations, that \textit{Swift} is not primed to do.}
    \label{fig:uvia_v_swift}
    
\end{figure}

\textbf{SNe Ia out-of-band signal:} SNe Ia are optically bright and may be $\sim$100$\times$ brighter in the optical band than in the UV at later times in the light curves (e.g., \cite{Mazzali+2014}). This anticipated brightness drives the need to reject optical light photons in the UV bands (red light suppression), and leads to the design choice to use multi-layer UV sensitive mirror coatings and detector filters that significantly block visible light photons. With a 2-mirror telescope design with narrow-band multi-layer coatings and metal dialectic detector filters, the out-of-band rejection of each UV channel as a function of wavelength is shown in Figure~\ref{fig:uvia_v_swift}~(right). UVIa's total red light suppression (all photons red of the designed bandpasses) is 7.5$\times$10$^{-5}$ (FUV) and 6.0$\times$10$^{-5}$ (NUV). Compared to \textit{Swift}-UVOT UV filters that most closely match UVIa's UV channels (UVM2 and UVW2)\cite{Breeveld+2010}, which have a minimum red light suppression of 2.6$\times$10$^{-3}$ (UVM2) and 5.2$\times$10$^{-3}$ (UVW2), UVIa suppresses 50 -- 100$\times$ more red photons.

We model the out-of-band contribution to the signal using the same representative SN Ia spectrum from SN 2011fe, which has full spectral coverage from 1150 -- 10000 {\AA} at one temporal point in its light curve\cite{Mazzali+2014}. When all optical light (3000 -- 10000 {\AA}) is integrated in each channel, the ratio of observed \textit{total} out-of-band to in-band flux detected by UVIa in the FUV, NUV, and $u$-band channels is 4\%, 3\%, and 0.5\%, respectively. 

Even at these levels, UVIa will calibrate out the out-of-band contribution. To accomplish this, we will characterize the integrated system response in the laboratory, ahead of on-orbit science operations. Using this and the concurrent optical observations from the ground, we can model the out-of-band contribution and remove it. We will further refine our red leak model and monitor the red leak for any significant variations or changes to the telescope behaviors (e.g., due to UV coating or MDF instabilities or degradation on-orbit over time) with frequent standard star observations.

\textbf{Detector noise:} The CIS120-LN detector noise characteristics are assumed uniform over all pixels  The CIS120-LN RN is reported to be $\sim$7 electrons/pixel (e/pix) by Teledyne-e2v and has been measured with a RN between 4 -- 6 e/pix during laboratory testing with spaceflight-ready electronics\cite{Otero+2021}. Dark current (DC) is intrinsic to silicon-based detectors and has well-calibrated curves as a function of temperature\cite{Jewell+2024}. At device temperatures achieved with passive cooling strategies, we expect a nominal DC rate of 0.01 e/pix/sec. This results in a total DC noise contribution less than the RN at exposure times $<$1000 seconds. By contrast, UVIa baselines an exposure time of 300 seconds per image -- \textcolor{black}{the total integration time for all three channels to meet the sensitivity requirements involves the stacking of shorter exposure times to build up the required total integration time.}. 
To ensure conservative estimates on performance, we assume an addition 30--50\% margin on both RN and DN to calculate UVIa's nominal sensitivity. This margin also accounts for increased noise from higher thermal conditions, camera electronics, and/or elevated noise from exposure to high radiation environments over time than anticipated.

\textbf{Earth geocoronal airglow:} Earth airglow affects UV observations in space, where recombination lines of hydrogen (Ly$\alpha$: $\lambda$1216) and oxygen ($\lambda$1304, 1356, 2471) produce strong emission features in the UV. We estimate the anticipated amount of stray light contamination from airglow, primarily from the lines listed above, in each of UVIa's channels. For each airglow line, we assume a conservative airglow line brightness by adapting twilight-level fluxes in the airglow features\cite{Leinert+1998,CruzAguirre2023}. We then use the full range of UVIa effective area curves, which cover 1000 -- 100000 {\AA}, to model the count rate of airglow lines detected by UVIa. The combined effects of the ML UV coatings, as well as MDF filters, allow for UVIa's UV channels to both be insensitive to optical light but also out-of-band UV light, which minimizes the harmful effects that geo-coronal airglow could have on a UV astronomical imaging telescope like UVIa. 

\textbf{Zodiacal light:} Zodiacal light provides an additional stray light background, particularly in the NUV and $u$-band channels, where scattered solar light off of small dust particles in the inner solar system and above the solar system plane\cite{Leinert+1998}. We estimate the contribution to background noise from zodiacal light in all UVIa channels by assuming a solar spectrum spectral energy distribution (SED) normalized to the worst-case brightness scenario, 17.8 V-mag/arcsec$^2$, across 1000 -- 10000 {\AA}\cite{Leinert+1998}. The solar spectrum used is from the National Renewable Energy Laboratory (NREL) and American Society for Testing and Materials (ASTM) E490 solar spectrum standard model, which is re-normalized to our maximum expected zodiacal light surface brightness and constitutes our zodiacal light spectrum. The zodiacal background has the strongest impact on the $u$-band channel, which is expected, given the solar spectrum approaches peak brightness near the $u$-band waveband. Contributions of zodiacal light in both UVIa UV channels is the lowest amount all potential noise sources (see Figure~\ref{fig:count_breakdown}).

\textbf{SNe Ia host galaxy:} To estimate the background contribution to the SNe Ia signal from its host galaxy, we assume a worst-case scenario contamination for a SNe Ia at the UVIa sensitivity limits in the UV (m = 21.5 AB) and that the host galaxy is unresolved. Assuming the majority of SNe Ia we would observe with UVIa will reside in galaxies forming stars, we use a galaxy SED template of a standard spiral arm galaxy generated with the \texttt{hyperz} software package\cite{Bolzonella+2000}. We re-normalize the galactic SED to 10$\times$ the peak absolute luminosity of a standard SNe Ia, as the total bolometric galactic luminosity is expected to be 10$\times$ brighter than the luminosity of the supernova\cite{Bolzonella+2000}, and shift the galaxy out to UVIa's maximum distance sensitivity (d $\sim$ 100 Mpc). We further assume that the host galaxy flux will be accounted for and subtracted out of the SNe Ia signal prior or posterior to the SNe Ia detection, such that the noise contribution left behind by the host galaxy in signal pixels is Poisson distributed. The host galaxy background signal, assuming a standard star forming galaxy. The SNe Ia host galaxy is on-par with the noise contributions from dark current. However as noted above, we take a conservative approach for the galactic contamination -- at a distance of 100 Mpc, galaxies larger than the Large Magellanic Cloud (e.g., $>$10 kpc) will be resolved at UVIa's spatial resolution ($<$1$^{\prime \prime}$) and plate scale (6$^{\prime \prime}$). This means that the host galaxy's flux will be spread over more pixels than the SNe Ia, bringing the noise contribution in the SNe Ia signal down by 1/pix$^2$.

\begin{figure}
    \centering
    \vspace{-0.5cm}
    \includegraphics[width=0.75\textwidth]
    {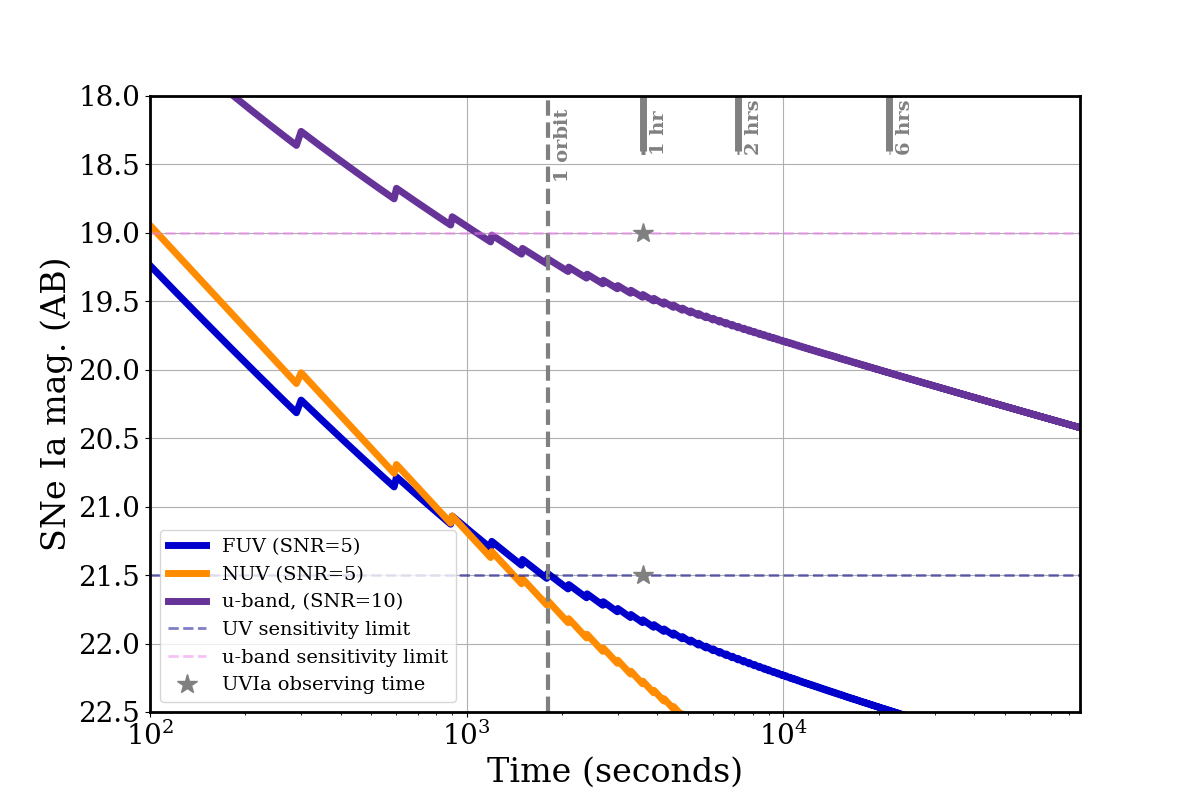}
    \caption{UVIa's limiting magnitude in the FUV (blue, SNR$\geq$5), NUV (orange, SNR$\geq$5) and $u$-band (purple, SNR$\geq$10) as a function of integration time. Each individual exposure per band is assumed to be 300 seconds. UVIa's science objectives accommodate up to 1 hour of integration time per channel, with is delineated with gray stars at the required sensitivity in the UV and $u$-band channels. However, as shown, UVIa's projected sensitivity is achieved within 2 -- 3 exposures in all channels (1200 -- 1800 seconds), which can be observed within an orbit, accounting for slew time, settling, and other observatory operations prior to science exposure. }
    
    \label{fig:snr}
\end{figure}

\textbf{Projected Signal-to-Noise Ratio (SNR):} 
Figure~\ref{fig:snr} shows the projected limiting magnitude in each UVIa band as a function of integration time, assuming the minimum SNR needed in each band is met. This exposure time calculation assumes that the sources of noise described above can be accounted for and corrected in the final data product. With UVIa's projected angular resolution, the resolution per pixel, and typical spacecraft jitter anticipated with commercial off-the-shelf CubeSat/SmallSat options ($\sim$10$^{\prime \prime}$), we coadd 2x2 pixels for the signal and all noise sources for the SNR estimation. We choose an exposure time of 300 seconds, which puts the signal in each band above the dominant noise source (usually RN) in each band (Figure~\ref{fig:count_breakdown}). UVIa's SNR requirements in all bands are met by 1800 seconds of integration time ($\sim$1 night-side orbit in Low Earth Orbit (LEO)), which gives $\sim$50\% margin on the targets that can be followed up by UVIa per day.

\subsection{UVIa Payload \& Spacecraft Requirements Overview}\label{sec:payload}
The UVIa payload is divided into the primary opto-mechanical system (FUV, NUV, and u-band telescopes and support structures) and supporting subsystems (electronics, thermal), which interface with the s/c bus.

\textbf{Payload Mechanical Requirements:} 
Figure~\ref{fig:payload_raytrace}(a) shows a top-level conceptual model of the UVIa payload design. The UVIa payload takes up a volume of 22cm x 25cm x 21.5cm, the minimum space needed to accommodate all three UVIa telescopes, an additional footprint of a co-aligned star tracking camera, and associated payload electronics. UVIa's spatial resolution requirements drive the need for stable spacecraft pointing during science exposures. Off-the-shelf CubeSat providers can supply up to 7$^{\prime \prime}$ RMS pointing stability (1-$\sigma$) but require additional on-board star tracking cameras to do so. The UVIa payload is designed to hold space for an additional star tracker, separate from the 3 telescopes, to ensure pointing stability is met. The minimum current best estimate (CBE) mass budget for the payload is $\sim$12 kg. 

UVIa's volume and mass profile fits within several commercially available 16-U CubeSat spacecraft options and can easily scale up to an ESPA-class SmallSat, depending on specific science objectives and requirements on UV sensitivity. As an example, Figure~\ref{fig:payload_Aeff}~(a) shows UVIa integrated into the University of Toronto Institute for Aerospace Studies (UTIAS) Space Flight Laboratory (SFL) JAEGAR 12/16U spacecraft.

\textbf{Payload Thermal Requirements:} The UVIa optical section is separated from the spacecraft bus and payload electronics to help thermally isolate the optics and detectors from payload and spacecraft components. Radiator plates to the deep space facing side of the spacecraft (same side as the telescope aperture doors) will direct heat generated in the payload to deep space and help passively cool the CMOS devices. The payload is designed to operate within a temperature range of 293$\pm$20 K, while each CMOS device is maintained at -~35$^{\circ}$C (to reach the maximum allowed DC) \textcolor{black}{via low-power PID thermal controllers}. Heaters will be installed along the optical bench, \textcolor{black}{including mounts and optics, to maintain an average temperature of 293 K} during on-orbit operations and science observations (e.g., during eclipse).

\textbf{Payload Power Requirements:} 
The UVIa payload power requirements are largely driven by the power consumption of the CMOS cameras, which includes on-chip processing (which handles programmable logic and signal conditioning to adjust bit resolution, gain, integration times, shutter speed, and image stacking options), as well as a low voltage differential signaling interface to the science instrument processor \textcolor{black}{that} provides control and image transmission capability. \textcolor{black}{Three CMOS sensors (one for each channel) and two sets of camera electronics (the FUV and NUV sensors share one set of camera electronics; see Figure~\ref{fig:payload_raytrace}) are baselined for the UVIa science payload.} Each CMOS sensor processor requires 450 mW of power during operations. The additional camera electronics themselves are allocated $\sim$5W during operation per camera. Additional power consumers, including a simple on-board computer to partially reduce and compress data for storage and downlink, heaters to keep optical and mechanical structures within acceptable temperature ranges, and power supply converters, are expected to require an additional $\sim$5W. During nominal operations, UVIa draws $\sim$10W power from the s/c, while at its absolute peak power usage, the \textcolor{black}{UVIa science payload requires} $<$20W. 

\textbf{Data generation:} 
UVIa's 1-hour integrated exposures per channel are achieved by stacking 12 shorter exposures (300 seconds) before processing. Each CIS120 device includes built-in processing and basic functionality, including the ability to integrate and stack multiple exposures before reading out to the on-board processor. Each device will read out to an on-board field programmable gate array (FPGA), which consists of programmable logic and signal conditioning to adjust bit resolution, gain, integration times, shutter speed, and image stacking options, among other controls. A low voltage differential signaling interface to the science instrument processor provides control and image transmission capability. Software on the science instrument processor will perform standard CMOS reduction and image compression for each stacked science image to prepare them for \textcolor{black}{downlink} to the ground.

With 14-bit resolution, each 2k$\times$2k full-frame image is $\sim$7.3 MB per CMOS camera. With an observing efficiency of 50\% (which includes anticipated anomalies, such as s/c tumbling, where potential observing time is lost) and assuming no compression of raw images, we anticipate a total daily data generation rate $\sim$250 MB/day (combined data generated for all of UVIa's channels). Payload housekeeping (e.g., power, temperature) adds $\leq$10 MB/day to the daily data rate (assuming 24 housekeeping analog channels that store 12 bits/channel with a sampling rate of 1 second, plus conversion to digital signal with overhead). The on-board computer (OBC) would baseline 64 GB of storage for the payload, or $\sim$2 months of data. 
To minimize downlink needs, we will stack frames into 1 hour composites on the OBC using standard techniques. Stacked images, along with housekeeping data, would be downlinked during ground passes. Before downlink, integrated images are compressed by $\geq$2 via a lossless compression algorithm (e.g., \texttt{fpack} \cite{fpack}). Further calibration/processing like flat fielding and photometry will be performed on the ground following well-developed automated CMOS data reduction tools developed by Las Cumbres Observatory (e.g., BANZAI\cite{McCullyBANZAI}). 

UVIa baselines an S-band receiver/transmitter, which provides 32 kbps uplink (primarily used to update UVIa's observing schedule with optimized observing strategies for newly discovered targets and detector calibration, s/c control, etc.) 
2--4 Mbps downlink speeds. UVIa baselines a minimum of 2 daily ground stations passes for data uplink and downlink. Assuming the minimum 2 Mbps downlink (S-band) with two 6-minute ground passes per day \textcolor{black}{per ground station}, UVIa's compressed daily data output would be downlinked with 34\% \textcolor{black}{growth} margin\cite{Karpati+2012}. 
Depending on UVIa's orbital inclination, a planned Mission Operations Center (MOC) at the University of Florida (UF) would serve as UVIa's primary ground station. The dedicated facility would meet the minimum two ground passes per day for full communications coverage.

\textbf{General Spacecraft Requirements:} 
UVIa's form factor, mass, power/thermal needs, pointing and stability requirements on image quality, and data downlink rates all drive mission-level requirements on s/c, which in turn limits viable, low-cost s/c options. At a minimum, the UVIa instrument fits within a little over a 10U volume and has a dry mass between 10 -- 15 kg. During normal operations, UVIa requires $\sim$20 W of power, not including active detector cooling (e.g., relying on radiative cooling alone). Compatibility with at least S-band telecomms is needed. One of the strictest requirements to achieve UVIa's science goals is on pointing stability ($<$10$^{\prime \prime}$ 1-$\sigma$ RMS) and, because of the 2-mirror \textcolor{black}{double off-axis} design \textcolor{black}{which degrades} image performance outside $\sim$1$^{\prime}$ on-axis focus \cite{CruzAguirreJATIS}, pointing knowledge. Most off-the-shelf from commercial Nano/Micro-Sat companies can meet most of UVIa's requirements, except in pointing. A few viable options exist, including the UTIAS SFL 12/16U JAEGAR bus, and the Blue Canyon Technologies (BCT) XB16U CubeSat.

\subsection{Nominal Orbit, Observing Strategy, \& Science Mission Profile}
UVIa's accomplishes its objectives with any Sun Synchronous Orbit (SSO) and any launch window with \textcolor{black}{a $\geq$500} km Low Earth Orbit (LEO)\cite{orbitm}, \textcolor{black}{though we note that a terminator (dawn-dusk) orbit would better maximize target observing/availability while optimizing solar array orientation (thus bus battery charging) and passive cooling via deep space-pointing radiators.} 
\textcolor{black}{However, considering mission operations at a conservative and common orbital configuration, we assume a noon-midnight orbit with} a maximum of 45 minutes of observing time (30 minutes in eclipse) and 45 minutes of Sun pointing the solar arrays to charge the batteries. 
Assuming Sun, Earth-limb, and Moon-limb keep-out zones of 65$^{\circ}$, 20$^{\circ}$, and 15$^{\circ}$, UVIa can access up to $\sim$70\% of the sky during the night side operations and up to $\sim$30\% on the day side of its orbit. 
The minimum telescope aperture size of the UVIa UV channels \textcolor{black}{(8 cm $\times$ 8 cm)} allows UVIa to observe SNe Ia out to $\sim$100 Mpc (down to 21.5 mag). At this distance, the expected rate of SNe Ia per year \textcolor{black}{is} 50 -- 100. The baseline science operations for UVIa in a 12U/16U CubeSat s/c is 18 months, which includes 50\% margin on the anticipated SNe Ia rates \textcolor{black}{(see Section~\ref{sec:rates})}. Increasing the aperture sizes lowers the required science mission time -- UVIa could accomplish its science objectives in a Pioneers-class mission profile within 1 year (including commissioning and closeout time). 

\textcolor{black}{Once UVIa reaches its nominal orbit, a one-time deployable aperture doors, which protects the payload during observatory integration and launch from contamination and moisture, release (Figure~\ref{fig:payload_raytrace}) and remain open for the remainder of the mission.} The mission profile includes frequent visits to standard calibration stars (e.g., white dwarfs) to monitor and adjust the instrument throughput response function over time, red leak performance over the course of the orbit, stability of UV mirror coatings and detector filters, and provide flux calibration for science targets. Calibration activities are especially important for assessing the on-orbit stability of flight-ready UV technologies primed for the Habitable World Observatory (HWO); see Cruz Aguirre\cite{CruzAguirreJATIS} for more details.

\subsubsection{Science Operations: From Target of Opportunity (ToO) Alerts to Light Curves}
UVIa takes advantage of emerging and demonstrated Time Domain and Multi-Messenger (TDAMM) infrastructure being developed by ground-based facilities (e.g. Las Cumbres Observatory). 

For UVIa, ToO alerts will be ingested from community brokers (e.g.~ANTARES \cite{antares}) and will be automatically ingested by a Target and Observation Manager (TOM) system similar to the Supernova Exchange (SNEx)\cite{Pellegrino2024} that has been developed for the Las Cumbres Observatory Global Supernova Project. Once the new targets are in the TOM, UVIa observations will be triggered and scheduled via the open-source Observatory Control System (OCS)\cite{NationSPIE}. The OCS is currently operating the 25 robotic telescopes of the ground-based Las Cumbres Observatory. This includes an automated scheduler, a data archive, and a portal for users to input observation requests for a given target. The current scheduler for Las Cumbres Observatory optimizes a metric that is a combination of observation length, schedule completeness, and scientific priority.  

During the UVIa development stage, these tools will be adapted for space-based systems. The primary development on the OCS will be adding the ability to account for orbital motion in the visibility of the telescope in the automated scheduler rather than the rotation of the earth for the ground-based observatories. We will also add slew time to the scheduler optimization weights, a feature from which both future ground-based and space-based observatories will benefit.

In parallel to triggering UVIa, the TOM system will automatically trigger ground-based facilities which are part of the Astrophysical Events Observatories Network (AEON\cite{RachelAEON}). These facilities will begin to observe the new supernova immediately, collecting photometry and spectroscopy. By automatically triggering both UVIa and ground-based follow-up observations, we have no additional lag waiting for a classification spectrum. Once a classification spectrum is obtained, the science team will vet the candidate to ensure it meets our scientific selection criteria. If the object does not, we will remove further observations from both ground-based observatories and UVIa. If the object does meet our selection criteria, UVIa will add it to our observing queue \textcolor{black}{via our TOM system. We will employ a dynamic observation cadence, sampling the light curve multiple times per day at early phases to fully capture and characterize early UV excesses. The cadences will be monitored from the ground by the science team via our TOM system. Any changes to the cadence will be reflected in the observing schedule we upload to the spacecraft at its regular uplinks.} At later phases, we will lower the cadence to every few days to match the speed of the light curve evolution. This strategy will allow us to build a high density light curve to capture the rapid evolution in the UV at early times while maximizing the number of objects we can follow at any given time. Ground-based facilities will typically monitor the supernova between a 1-day and 3-day cadence depending on the phase of the light curve and how rapidly the supernova is evolving in the optical. 

The cadence is controlled by the TOM system, trying to maintain a fixed cadence, but reacting to missed observations (e.g. bad weather on the ground). If an observation is weathered out for example, the TOM resubmits an observation request immediately without waiting until the next scheduled observation. This maximizes the uniformity of the sampling of light curves in the non-ideal case when some observations cannot be obtained.

Once the supernova has dropped below the detectability thresholds of the instruments of interest, we will remove the object from the scheduler. We anticipate following a similar workflow to how the GSP monitors which supernovae are being observed, rotating monitoring duties between core team members.

In summary, our science operations plan is as follows:
\begin{enumerate}
    \item Large all-sky surveys (e.g., ZTF, LSST) will alert the community about transient candidates through public brokers like ANATARES \cite{antares} and ALERCE \cite{alerce1,alerce2,alerce3,alerce4}.
    \item UVIa's TOM monitors the stream, automatically adding targets and triggering observations on UVIa of any SNe discovered in galaxies within 100 Mpc that have recent non-detections, suggesting the objects are young.
    \item In parallel, dedicated ground-based programs will be triggered to classify the new SN (e.~g. Las Cumbres Observatory's FLOYDS spectrograph (R$\approx500$ from 3400 -- 10000 \AA\ \cite{Brown13})  
    \item Simultaneously, photometric observations will be triggered to build complementary optical (ugriz) light curves of the supernova. 
    \item Once classified, the science team will vet the objects, removing objects that do not meet the science needs.
    \item For objects that remain in the queue, the scheduler will continue to include observations on UVIa until the full light curve is collected. 
\end{enumerate}

\section{UVIa in the Landscape of Current and Upcoming Missions}
UVIa stands to produce an unprecedented UV dataset of SNe Ia. These lightcurves will have a significant legacy value for both upcoming missions and for the theoretical modeling of the SN Ia explosion mechanism. 

\textbf{\textit{Swift:}} UVIa builds on the foundations laid by Swift's supernova program, adding significant red suppression, removing a key systematic in SN Ia analysis (see Section~\ref{sec:performance} for a more detailed discussion). UVIa's dedicated monitoring campaign also ensures that we obtain more uniformly sampled and more complete SN Ia lightcurves. 

\textbf{ULTRASAT:} UVIa and ULTRASAT will both observe SNe Ia, but will produce different populations. UVIa is designed to be an all-sky, reactive follow-up facility. This allows us to observe a large fraction of SNe Ia in the local volume. ULTRASAT adopts a narrow-deep approach. Each mission will collect a similar number of objects (UVIa will collect the same number of objects in half the time), and the majority of ULTRASAT's objects will be more distant, limiting the connection to more nearby rungs of the distance ladder. UVIa's sample is more optimized for calibrating the SN Ia sample for cosmology. Combining these complementary samples will place the strongest current constraints on the presence of UV excesses in SNe Ia to date. UVIa's lightcurves have the added benefit of including simultaneous optical ($u$-band) and FUV observations rather than ULTRASAT's single NUV band. These multi-wavelength data will be invaluable to model the full SED for cosmological analyses and to provide constraints on explosion models via temperature evolution (see Section \ref{sec:science}). 

\textbf{UVEX:} \textcolor{black}{UVIa can cover vital discovery space ahead of/in conjunction with missions like UVEX, allowing larger, more powerful facilities to optimize their response and observing cadence on targets of interest and maximizing the execution of science programs over their limited mission lifetimes. While SNe Ia are not one of the UVEX mission's primary objectives, UVIa's UV observations of SNe Ia would provide key motivation for SNe Ia-focused investigations as part of UVEX's Community Target of Opportunity (ToO) program.} Currently, the FUV is a nearly completely unexplored wavelength region for SNe Ia. UVIa will provide the first sample of FUV observations of SNe Ia, which can be used to inform the Community ToO Science Objective currently in place by UVEX, providing the necessary context and brightness expectations for enabling these ToOs for SNe Ia programs. UVIa paves the way for UVEX to explore and discover spectroscopic information about SNe Ia that would further uncover their origins and explosion mechanisms. \textcolor{black}{However, should UVIa operate after the completion of UVEX, the combinations of serendipitous discovery of SN Ia through UVEX's all-sky archive and SN Ia-oriented ToO programs would only benefit UVIa's survey and sample statistics for cosmological templates and progenitor population statistics, much like ULTRASAT's NUV sample.}

UVIa stands to make an outsized contribution to SNe Ia cosmology for its cost cap, providing template lightcurves for upcoming cosmological surveys like those of Roman and Rubin. The combination of recent technology advancements and the landscape of upcoming missions makes this the opportune time for UVIa to collect its legacy SN Ia UV dataset. 

\section{Future Direction}
While we have presented UVIa in the context of SNe Ia physics and cosmology, UVIa acts as a prototype for larger-scale, general TDAMM science. Given the comparably low cost of the UVIa CubeSat platform, this technology can scale to produce a network of TDAMM follow-up telescopes to form a network reminiscent of the one produced by Las Cumbres Observatory on the ground. The scheduling and operations of UVIa also generalize to other wavebands like optical or infrared and to larger platforms like SmallSats.

TDAMM science is and will likely remain a priority in astronomical research. To ensure success in this area, we need to continue to evolve our observational strategy, especially for space-based missions. Las Cumbres Observatory (and other telescope networks with similar aims) have proven to be important for follow-up observations of a wide variety of time-variable events. Discovery by surveys is only the first step. For a physical understanding of these events, we need coverage of the SED beyond single-band photometry. These facilities need to be able to respond quickly to observe these events to capture critical parts of their evolution. UVIa acts as a pathfinder for this type of follow-up astronomy. The long-term goal of this project is to produce a relatively inexpensive platform (through programs like NASA ROSES APRA and Astrophysics Pioneers) for space-based UV/O(/IR) TDAMM astronomy that will scale to the needs of current and upcoming surveys, which the team plans to propose for to support critical transient astronomy infrastructure.

\subsection* {Code, Data, and Materials Availability} 
The code and data used in this work are available upon request to the authors.

\subsection* {Acknowledgments}
K.H. acknowledges support through the NASA Roman Technology Fellowship (80NSSC24K0471). 
Time-domain research by the University of Arizona team and D.J.S.\ is supported by National Science Foundation (NSF) grants 2108032, 2308181, 2407566, and 2432036 and the Heising-Simons Foundation under grant \#2020-1864. 
The research was carried out in part at the Jet Propulsion Laboratory, California Institute of Technology, under a contract with the National Aeronautics and Space Administration (80NM0018D0004). The authors would like to thank of the two reviewers of this manuscript, who provided thoughtful feedback that helped to make this manuscript clearer.

\subsection* {Disclosures}
The authors declare there are no financial interests, commercial affiliations, or other potential conflicts of interest that have influenced the objectivity of this research or the writing of this paper.

\bibliography{report}   
\bibliographystyle{spiejour}

\vspace{2ex}\noindent\textbf{Keri Hoadley} is an Associate Professor in the Department of Astronomy at the University of Florida. She received her BS in Astronomy/Astrophysics and Mathematical Sciences from Florida Institute of Technology in 2011 and her MS and PhD degrees in Astrophysics from the University of Colorado in 2014 and 2017, respectively. She received the David \& Ellen Lee Postdoctoral Fellowship in Experimental Physics at California Institute of Technology in 2018 and was named a NASA Roman Technology Fellow in 2021. Her current research interests include using ultraviolet light to understand how structures in the universe form and evolve, and developing ultraviolet mission concepts and technology maturation for astronomical applications. She is a member of SPIE and is co-chair of the ``UV, X-Ray, and Gamma-Ray Space Instrumentation for Astronomy'' conference for SPIE Optics and Photonics.

\vspace{1ex}\noindent\textbf{Curtis McCully} is a Senior Astrodata Scientist at Las Cumbres Observatory. He completed his Ph.D. at Rutgers, the State University of New Jersey, in 2014 on observations of peculiar thermonuclear supernovae and gravitational lensing theory. He took a postdoc fellowship at Las Cumbres Observatory, working on a variety of explosive transients, including type Ia supernovae. He is a leader in the gravitational-wave follow-up group, which co-discovered the first confirmed kilonova in 2017. In his current position, McCully bridges the gap between scientists and software engineers, opening new capabilities in time-domain astronomy by developing data reduction pipelines and building interfaces for scientists and their data.

\vspace{1ex}
\noindent Biographies and photographs of the other authors are not available.

\listoffigures
\listoftables

\end{spacing}
\end{document}